\documentclass[10pt,fleqn]{article}

\usepackage{amsmath,amsfonts,amssymb,latexsym,cite}
\usepackage{graphicx}
\usepackage{color}

\usepackage{amsfonts,amssymb,cite}
\usepackage{graphicx}

\def\noi{\noindent}
\def\jnumber#1#2{\thispagestyle{empty} \noi\unitlength=1mm
    	\begin{picture}(178,10)
            \put(177,15){\llap{\large\it Grav. Cosmol. No.\,#1, #2}}
                    \end{picture}}

\newcommand{\Title}[1]{\noi {{\Large\bf #1}}\\[1ex]}

\def\Aunames#1{\noi{\bf #1}}
\def\au#1{${}^{#1}$}
\def\Addresses#1{\medskip\noi \protect
	\begin{description}\itemsep -3pt {\it #1} \end{description}}
\def\adr#1#2{\item[${}^{#1}$]{\it #2}}

\def\email#1#2{\footnotetext[#1]{e-mail: #2}\addtocounter{footnote}{1}}

\def\nqq{\hspace*{-2em}}
\def\nhq{\hspace*{-0.5em}}

\def\Jl#1#2{#1 {\bf #2},\ }

\def\ApJ#1 {\Jl{Astroph. J.}{#1}}
\def\CQG#1 {\Jl{Class. Quantum Grav.}{#1}}
\def\DAN#1 {\Jl{Dokl. AN SSSR}{#1}}
\def\GC#1 {\Jl{Grav. Cosmol.}{#1}}
\def\GRG#1 {\Jl{Gen. Rel. Grav.}{#1}}
\def\JETF#1 {\Jl{Zh. Eksp. Teor. Fiz.}{#1}}
\def\JETP#1 {\Jl{Sov. Phys. JETP}{#1}}
\def\JHEP#1 {\Jl{JHEP}{#1}}
\def\JMP#1 {\Jl{J. Math. Phys.}{#1}}
\def\NPB#1 {\Jl{Nucl. Phys. B}{#1}}
\def\NP#1 {\Jl{Nucl. Phys.}{#1}}
\def\PLA#1 {\Jl{Phys. Lett. A}{#1}}
\def\PLB#1 {\Jl{Phys. Lett. B}{#1}}
\def\PRD#1 {\Jl{Phys. Rev. D}{#1}}
\def\PRL#1 {\Jl{Phys. Rev. Lett.}{#1}}

\def\al{&\nhq}
\def\lal{&&\nqq {}}
\def\eq{Eq.\,}

\def\beq{\begin{equation}}
\def\eeq{\end{equation}}
\def\bear{\begin{eqnarray}}
\def\bearr{\begin{eqnarray} \lal}
\def\ear{\end{eqnarray}}
\def\earn{\nonumber \end{eqnarray}}

\def\eqv{\al \equiv \al}

\begin{document}
\onecolumn
\jnumber{x}{2021}

\Title{Gravitational baryogenesis of cosmological constant dominated universe}

\Aunames{Sudhir Kumar Srivastava\au{a,1}, Ahamad T Ali\au{b,2}, Anil Kumar Yadav\au{c,3}} 

\Addresses{
\adr a {\small Department of Mathematics, Janata Janardan Inter College, Gandhi Nagar, Ghazipur 233225, India}
\adr b {\small Department of Mathematics, Faculty of Science King Abdul Aziz University, PO Box 80203
Jeddah, 21589, Saudi Arabia}
\adr c {\small Department of Physics, United College of Engineering and Research, Greater Noida - 201306, India}}

{\it Received {}}
\abstract
{In this paper, we have studied the gravitational baryogenesis of isotropic and homogeneous universe in the frame-work of general relativity. We investigate an exact and new solution of Einstein's field equations for FRW metric. Our solution represents a transitioning model of the universe which was expanding in decelerated mode and it transit in accelerated mode after dominance of cosmological constant $\Lambda$. We observe that gravitational baryogenesis occurs in the derived universe and derived baryon entropy ratio is in good agreement with its observational value.}\\

\textbf{Kewwords:} Gravitational baryogenesis, General relativity and Cosmological constant.\\


\email 1 {drsudhirgzp@gmail.com}
\email 2 {ahmadtawfik95@gmail.com}
\email 3 {abanilyadav@yahoo.co.in}

{ 
\def\mn{_{\mu\nu}}
\def\MN{^{\mu\nu}}
\def\mN{_\mu^\nu}
\def\nM{_\nu^\mu}
\def\cK{{\cal K}}
\def\cV{{\cal V}}
\def\eqv{\al \equiv \al}
\def\kappa{\varkappa}

\def\wt{\widetilde}
\def\tg{{\wt g}}
\def\tR{{\wt R}}

\def\M{{\mathbb M}}
\def\N{{\mathbb N}}
\def\R{{\mathbb R}}
\def\S{{\mathbb S}}
\def\V{{\mathbb V}}
\def\oR{{\overline R}}

\def\rf{\eqref}
\def\eqn{\eq\eqref}

\def\bh{black hole}
\def\bhs{black holes}
\def\Swz{Schwarz\-schild}

\vspace{2mm}
\section{Introduction}
The cosmological observations have confirmed that the universe is in accelerated mode of expansion at present epoch \cite{Riess98,Perlmutter99}. It has been cited that an unknown form of energy with negative pressure is the reason for the accelerated expansion of the universe. Though the exact form of this energy is still unknown, but the recent observational results indicate that this unkwown energy has occupied about 68\% of the total energy budget of the universe \cite{Ade/2016}. This mysterious form of energy is known as dark energy and it was less effective in early stage of evolution of the universe but it dominates the present universe. Since it does not interact with the baryonic matter, hence it is hard to detect the dark energy. Theoretically, this can be studied with the cosmic expansion history $H(z)$ and the growth rate of cosmic large scale structure $f_g(z)$ \cite{Ma99}. Several theoretical models based on dark energy showing acceleration expansion of the universe have been proposed in last two decades \cite{Mishra15,Mishra17,Mishra18a,Yadav/2011,Yadav/2016,Yadav/2012epjp,Kumar/2011,Bennett/2013,Akarsu/2019prd,Amirhashchi/2019a,Amirhashchi/2019}. In particular, Akarsu et al. \cite{Akarsu/2019prd} have considered the simplest anisotropic generalization, as a correction, to the cosmological constant cold dark matter ($\Lambda$CDM) model, by replacing the spatially flat Robertson-Walker metric by the Bianchi type-I metric, which brings in a new term anisotropic density in the average expansion rate $H(a)$ of the Universe. In Ref. \cite{Amirhashchi/2019a}, Amirhashchi and Amirhashchi have studied Bianchi type I cosmological model and constrained this model with type Ia Supernova and H(z) Data. Later on, Amirhashchi and Amirhashchi \cite{Amirhashchi/2019} have studied XCDM and $\Lambda$CDM model using Gaussian processes. The most suitable and prominent candidate of dark energy is the cosmological constant $\Lambda$. Apart from fine tuning and cosmic coincidence problems, the $\Lambda$CDM model is the promising cosmological model to describe dynamics of the universe at present epoch \cite{Goswami/2016,Goswami/2016ijtp,Kumar/2015}.

The Big-Bang Nucleosynthesis (BBN) and Cosmic Microwave Background (CMB) predicts that the universe contains an excess of matter over antimatter \cite{Burles/2001,Bennett/2003}. Ade et al \cite{Ade/2016} have obtained observational constraints on baryon-entropy ratio as $\eta_{B} \leq 9\times 10^{-11}$. Baryogenesis is a theoretical process which fall out in the early stage of evolution of the universe. However, in modern cosmology, we believe that all particles burst into cosmos, follows same law of physics. Thus, the production of equal amount of matter and antimatter
must lead to zero baryon number in the universe. The interactions beyond the standard models have been studied to decode the matter and antimatter engima in Refs \cite{Stewart/1996,Yamada/2016,Akita/2017,kolb/1996}. In 1981, Dolgov had investigated thermal baryogenesis for black hole evaporation \cite{Dolgov/1981}. However, to generate baryon asymmetry, there are mainly three conditions i) violation of net baryon number ii) violation of Charge and Charge-Parity symmetry and iii) interactions beyond thermal equilibrium \cite{Sakharov/1967}. These conditions are known as Sakharov's criterion for baryon asymmetry. To satisfy the latter two conditions, the conventional approach has been to innovate interactions that violate Charge and Charge-Parity symmetry in vacuum and a period when the universe is evolving beyond the thermal equilibrium.

The violation of baryon asymmetry has been also reported in Refs \cite{Davoudiasl/2004,Cohen/1987}. In Refs. \cite{Davoudiasl/2004}, the authors have investigated that baryon asymmetry exist in an expanding universe with thermal equilibrium and without Charge-Parity symmetry. In the recent past, the concept of baryogenesis has been experimentally verified which trigger cosmologist to think in this direction. Some pioneer works on gravitational baryogenesis are given in Refs. \cite{Ramos/2017,Odintsov/2016,Oikonomou/2016,Baffou/2019,Bento/2005}. Recently, Bhattacharjee and Sahoo \cite{Bhattacharjee/2020a,Bhattacharjee/2020b} have studied gravitational baryogenesis in $f(Q,T)$ and $f(R,T)$ gravity respectively.

In this paper, we have investigated gravitational baryogenesis of isotropic and homogeneous universe in the frame-work of general relativity. We also obtain an exact and stable solution of Einstein's fields equations that generate baryon asymmetry in accelerating universe. Note that our work is is altogether different from Akarsu et al \cite{Akarsu/2019prd}. In fact, Akarsu et al \cite{Akarsu/2019prd} have investigated a Bianchi type I cosmological model that fit well with observations. The shape of the paper is as follows: Section 1 is introductory in nature. In section 2, we have discussed the model and its physical properties. We have examined the stability of solution in section 3. Section 4 deals with the occurrence of gravitational baryogenesis in the derived universe. Finally the conclusion is given in section 5.\\
\section{The model and its physical properties}
The isotropic and homogeneous gravitational field is read as
\begin{equation}
\label{metric}
ds^{2}\,=\,dt^{2}-a^{2}(t)(dx^{2}+dy^{2}+dz^{2}).
\end{equation}
where $a(t)$ is the scale factor.\\
The Einstein's field equation with cosmological constant $(\Lambda)$ is given by
\begin{equation}
\label{ef}
R_{ij}-\dfrac{1}{2}\,R\,g_{ij}-\Lambda\,g_{ij}\,=\,8\,\pi\,G\,T_{ij}.
\end{equation}
where $R$ is the Ricci scalar and other symbols have their usual meaning.

The energy-momentum tensor $(T_{ij})$ of perfect fluid is read as
\begin{equation}
\label{ef-1}
T_{ij}\,=\,(\rho+p)\,v_{i}\,v_{j}-p\,g_{ij}.
\end{equation}
where $v^{i}$ is four velocity vector satisfying $v^{i}\,v_{i}\,=\,1$. 

In equation (\ref{ef-1}), p and $\rho$ are the isotropic pressure and energy density of the fluid under consideration.\\

Solving (\ref{ef}) with space-time (\ref{metric}), we obtain the following system of equations
\begin{equation}
\label{ef-2}
\dfrac{2\,\ddot{a}}{a}+\dfrac{\dot{a}^{2}}{a^{2}}\,=\,-8\,\pi\,G\,p+\Lambda,
\end{equation}
\begin{equation}
\label{ef-3}
\dfrac{3\,\dot{a}^{2}}{a^{2}}\,=\,8\,\pi\,G\,\rho+\Lambda.
\end{equation}
The well known equation of state for perfect fluid is given by
\begin{equation}
\label{es}
p\,=\,\omega\,\rho.
\end{equation}
\begin{figure}[ht!]
\centering
\includegraphics[width=14cm,height=12cm,angle=0]{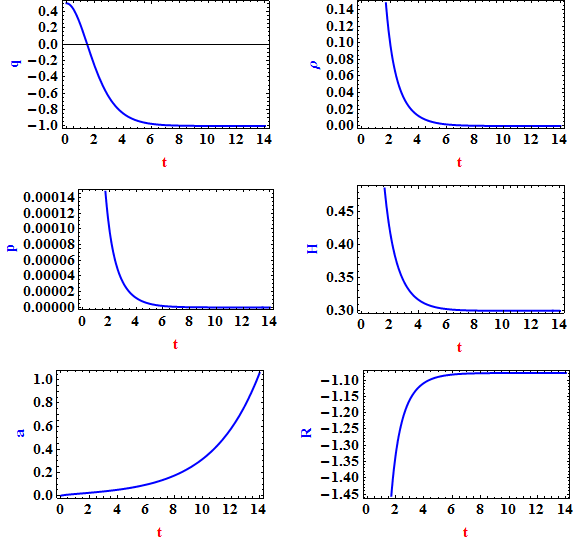}
\caption{Dynamics of model parameters versus time for $m = 0.025$, $\Lambda_{0} = 0.45$, $\omega = 0.001$ and $t_{0} = 0.002$.}
\end{figure}
where $\omega$ is known as equation of state parameter.\\
Using (\ref{ef-2}) and (\ref{ef-3}), the equation (\ref{es}) becomes:
\begin{equation}
\label{es-1}
\dfrac{2\,\ddot{a}}{a}+(1+3\,\omega)\,\left(\dfrac{\dot{a}}{a}\right)^2\,=\,(1+\omega)\,\Lambda.
\end{equation}
The general solution of equation (\ref{es-1}) is obtained as
\begin{equation}\label{a}
a(t)\,=\,m\,\sinh^{\dfrac{2}{3\,(1+\omega)}}\left(\Lambda_0\,(1+\omega)\,t+t_0\right).
\end{equation}
where $m$ and $t_{0}$ are constants of integration while $\Lambda_{0}\,=\,\sqrt{\dfrac{3\,\Lambda}{4}}$.\\
Using equation (\ref{a}) in equations (\ref{ef-2}) and (\ref{ef-3}), the expressions for energy density and pressure are obtained as
\begin{equation}
\label{rho}
\rho(t)\,=\,\dfrac{\Lambda_0^2\,\mathrm{\mathrm{csch}}^2\left[\Lambda_0\,(1+\omega)\,t+t_0\right]}{6\,\pi\,G}.
\end{equation}
\begin{equation}\label{p}
p(t)\,=\,\dfrac{\omega\,\Lambda_0^2\,\mathrm{csch}^2\left[\Lambda_0\,(1+\omega)\,t+t_0\right]}{6\,\pi\,G}.
\end{equation}
Here, $\mathrm{csch\left[\Lambda_0\,(1+\omega)\,t+t_0\right]}$ stands for $\mathrm{cosech\left[\Lambda_0\,(1+\omega)\,t+t_0\right]}$.

The deceleration parameter $q$ and Hubble parameter $H$ are read as
\begin{equation}\label{u2-DP-1}
q\,=\,-\dfrac{a\,\ddot{a}}{\dot{a}^{2}}\,=\,-1+\dfrac{3\,(1+\omega)}{2}\,\mathrm{sech}^2\left[\Lambda_0\,(1+\omega)\,t+t_0\right].
\end{equation}
\begin{equation}\label{u2-DP-1}
H\,=\,\dfrac{\dot{a}}{a}\,=\,\dfrac{2\,\Lambda_0}{3}\,\mathrm{coth}\left[\Lambda_0\,(1+\omega)\,t+t_0\right].
\end{equation}

The behaviour of kinematical and physical parameters of derived model are graphed in Fig. 1. From upper left panel of Fig. 1, we observe that the universe was expanding with decelerating phase in its initial epoch and it turns into accelerating phase with dominance of cosmological constant. This reflects that the derived universe represents a model of transitioning universe from matter dominated era cosmological constant dominated era.\\
\begin{figure}[ht!]
\centering
\includegraphics[width=6cm,height=6cm,angle=0]{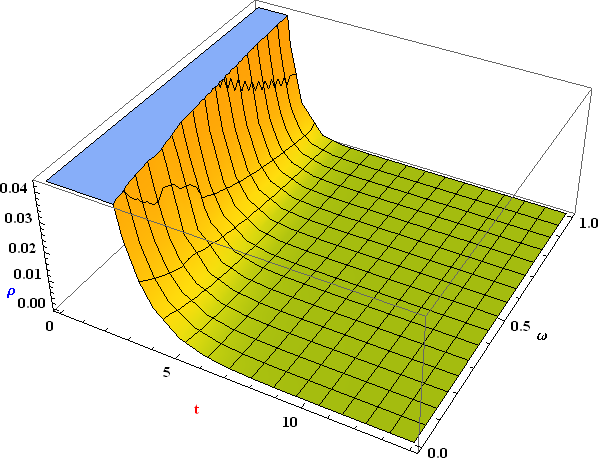}
\includegraphics[width=6cm,height=6cm,angle=0]{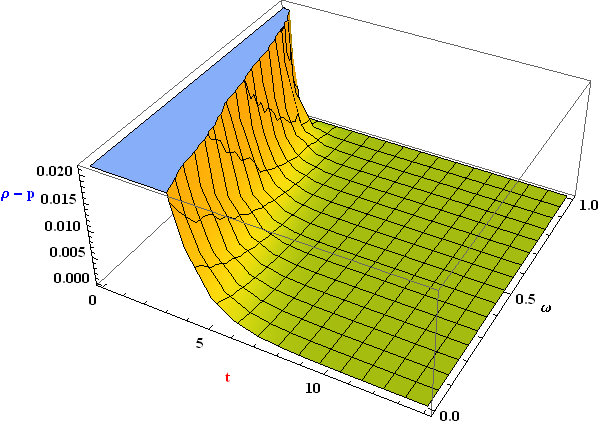}\\
\includegraphics[width=6cm,height=6cm,angle=0]{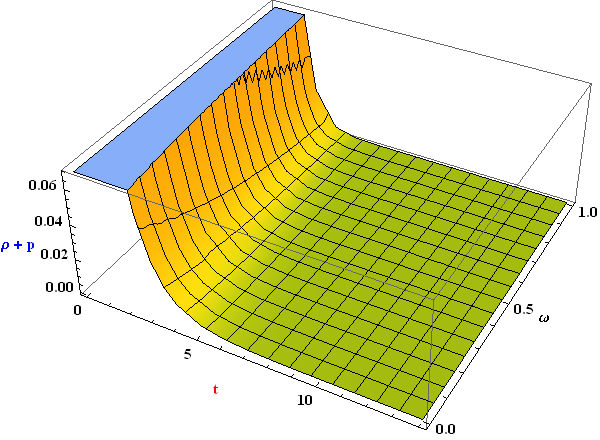}
\includegraphics[width=6cm,height=6cm,angle=0]{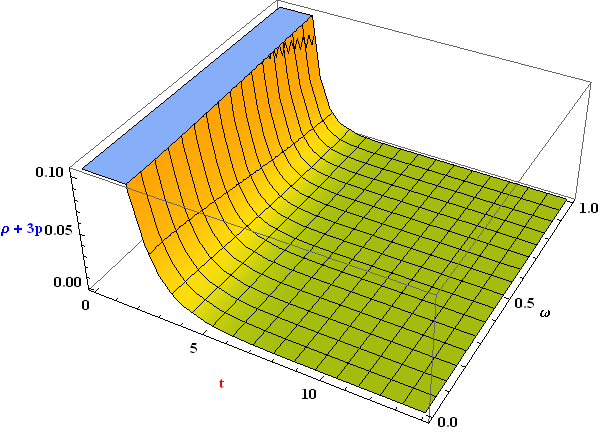}
\caption{Validation of energy conditions for $\Lambda_{0} = 0.45$ and $t_{0} = 0.002$.}
\end{figure}

The energy momentum tensor is usually required to satisfy four different energy conditions \cite{Wald/1984,Singla/2020}.
The four different types of energy conditions are described as follows.\\
i) Null energy condition (NEC): $\rho \geq 0$\\
ii) Weak energy condition (WEC): $\rho + p \geq 0$\\
iii) Dominant energy condition (DEC): $\rho - p \geq 0$\\
iv) Strong energy condition (SEC): $\rho + 3p \geq 0$\\
The above energy conditions are not independent. One can observe that if (i) DEC holds,
also the WEC holds (ii) if WEC holds, also NEC holds (iii) if SEC holds, also NEC
holds. From these relations, we can conclude that WEC and NEC are the important
energy conditions as their violation leads to the violation of other energy conditions. The validation of energy conditions in our derived model are shown in Fig. 2.\\
\section{Stability of solution}
In this section, we check the stability of background solution with respect to the time dependent isotropic perturbation of the metric as following.
\begin{equation}
\label{stability1}
a_{i}\,\rightarrow\,a_{Bi}+\delta a_{i} (t)\,=\,a_{Bi}\,(1+\delta b_{i}(t))
\end{equation}
where $i = 1, 2, 3$ along spatial directions $x$, $y$ and $z$. For FRW metric, $a_{1} = a_{2} = a_{3}$.\\
With reference to equation (\ref{stability1}), the perturbations of volume scalar, directional expansion scalar and mean expansion scalar are read as
\begin{equation}
\label{stability2}
V\,\rightarrow\,V_{B}+V_{B}\,\sum_{i}\delta b_{i},\,\,\,\,\,\theta_{i}\,\rightarrow\,\theta_{Bi}+\sum_{i}\delta b_{i},\,\,\,\,\,
\theta\,\rightarrow\,\theta_{B}+\dfrac{1}{3}\,\sum_{i}\delta b_{i}.
\end{equation}
The metric perturbation $\delta b_{i}$ satifies the following equations \cite{Saha/2012,Yadav/2019,Sharma/2020,Sharma/2018}.
\begin{equation}
\label{st3}
\sum_{i}\delta\ddot{b}_{i}+2\,\sum\theta_{Bi}\,\delta\dot{b}_{i}\,=\,0.
\end{equation}
\begin{equation}
\label{st4}
\delta\ddot{b}_{i}+\dfrac{\dot{V}_{B}}{V_{B}}\delta\dot{b}_{i}+\theta_{Bi}\,\sum_{j}\delta\dot{b}_{j}\,=\,0.
\end{equation}
\begin{equation}
\label{st5}
\sum\delta\dot{b}_{i} = 0.
\end{equation}
\begin{figure}[ht!]
\centering
\includegraphics[width=8cm,height=6cm,angle=0]{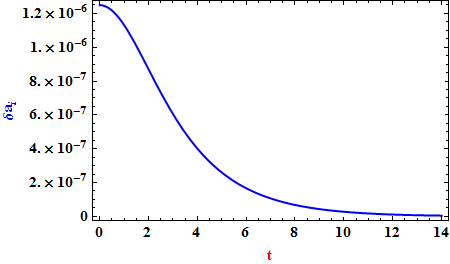}
\caption{The behaviour of $\delta a_{i}$ versus time for $\epsilon_{0} = 0.0003$, $\epsilon_{1} = 0.005$, $m = 0.025$, $\Lambda_{0} = 0.45$, $\omega = 0.001$ and $t_{0} = 0.002$.}
\end{figure}
Equations (\ref{st3})$-$(\ref{st5}) lead to
\begin{equation}
\label{st6}
\dfrac{\delta\ddot{b}_{i}}{\delta\dot{b}_{i}}+\dfrac{\dot{V}_{B}}{V_{B}}\,=\,0,
\end{equation}
where $V_{B}$ is the background spatial volume and it is obtained as
\begin{equation}
\label{st7}
V_{B}\,=\,m^{3}\,\sinh^{\dfrac{2}{(1+\omega)}}\left[\Lambda_0\,(1+\omega)\,t+t_0\right].
\end{equation}
Integrating the equations (\ref{st6}), we have
\begin{equation}
\label{st6-0}
\delta\dot{b}_{i}\,=\,\dfrac{\Lambda_0\,(1+\omega)\,\epsilon_{0}\,m^3}{V_B}\,=\,
\Lambda_0\,(1+\omega)\,\epsilon_0\,\sinh^{-\dfrac{2}{(1+\omega)}}\left[\Lambda_0\,(1+\omega)\,t+t_0\right],
\end{equation}
where $\epsilon_0$ is a constant of integration.\\
Integrating equation (\ref{st6-0}), we obtain
\begin{equation}
\label{st8}
\begin{array}{ll}
\delta b_{i}\,=\,\big(-1\big)^{\dfrac{3+\omega}{2\,(1+\omega)}}\,\epsilon_0\,\cosh\left(\,\Lambda_0\,(1+\omega)\,t+t_0\right)\\
\\
\,\,\,\,\,\,\,\,\,\,\,\,\,\,\,\,\,\,
\times\,\mathbf{F}\left[\dfrac{1}{2},\,\dfrac{1}{2}+\dfrac{1}{1+\omega};\,\dfrac{3}{2};\,
\cosh^2\left(\,\Lambda_0\,(1+\omega)\,t+t_0\right)\right]+\epsilon_1,
\end{array}
\end{equation}
where $\epsilon_1$ is a constant of integration and $\mathbf{F}$ is Gauss Hypergeometric function defined as
$$
\mathbf{F}\left[a,\,b;\,c;\,z\right]\,=\,\dfrac{\Gamma(c)}{\Gamma(a)\,\Gamma(b)}\,\sum_{k\,=\,0}^{\infty}\dfrac{\Gamma(a+k)\,\Gamma(b+k)\,z^k}{k!\,\Gamma(c+k)}
$$
$$
\,=\,_2\mathbf{F}_1\left[a,\,b;\,c;\,z\right]\,=\,\sum_{k\,=\,0}^{\infty}\dfrac{(a)_k\,(b)_k\,z^k}{k!\,(c)_k}
$$
$$
\,=\,1+\dfrac{a\,b\,z}{c}+\dfrac{a\,(1+a)\,b\,(1+b)\,z^2}{2\,c\,(1+c)}+\dfrac{a\,(1+a)\,(2+a)\,b\,(1+b)\,(2+b)\,z^3}{6\,c\,(1+c)\,(2+c)}+....
$$
Thus, the actual fluctuations $\delta a_{i} = a_{Bi}\delta b_{i}$ is obtained as
\begin{equation}
\label{st9}
\begin{array}{ll}
\delta a_{i}\,=\,m\,\sinh^{\dfrac{2}{3\,(1+\omega)}}\left(\Lambda_0\,(1+\omega)\,t+t_0\right)\Bigg(
\big(-1\big)^{\dfrac{3+\omega}{2\,(1+\omega)}}\,\epsilon_0\,\cosh\left(\,\Lambda_0\,(1+\omega)\,t+t_0\right)\\
\\
\,\,\,\,\,\,\,\,\,\,\,\,\,\,\,\,\,\,
\times\,\mathbf{F}\left[\dfrac{1}{2},\,\dfrac{1}{2}+\dfrac{1}{1+\omega};\,\dfrac{3}{2};\,
\cosh^2\left(\,\Lambda_0\,(1+\omega)\,t+t_0\right)\right]+\epsilon_1\Bigg).
\end{array}
\end{equation}
The behaviour of actual fluctuations $\delta a_{i}$ versus time is depicted in Figure 3. We observe that $\delta a_{i}$ starts with small positive value at $t\,=\,0$ and it decreases with evolution of the universe. At late time, $\delta a_{i} \rightarrow 0$, which indicates that the obtained solution is stable against the perturbation of gravitational field.
\section{Gravitational baryogenesis}
The baryon asymmetry factor which determines the baryogenesis is given by
\begin{equation}
\label{b-1}
\eta_{B}\,=\,\dfrac{n_{B}-\bar{n}_{B}}{s}.
\end{equation}
where $n_{B}$ and $\bar{n}_{B}$ denote the baryon and anti-baryon numbers respectively and $s$ is the entropy of the universe.

Following Ade et al \cite{Ade/2016}, the CMB and Big Bang Nucleosynthesis observational constraints on baryon asymmetry factor as $\eta_{B} \leq 9\times 10^{-11}$. In 2004, Davoudiasl et al \cite{Davo/2004} have introduced a mechanism to generate baryon asymmetry due to interaction between baryon number current density $J^{\mu}$ and $\dot{R}$, as
$$\dfrac{\pm 1}{M_{\star}^{2}}\,\int J^{\mu}\,\sqrt{-g}\,(\partial_{\mu}R)\,d^{4}x;$$
where $M_{\star}$ is cutoff scale of energy in effective theory \cite{Nozari/2018}.

We assume $\dot{R}\,=\,\mu_{B}{M_{\star}^{2}}$ is the total baryonic chemical potential and $g_{b}$ is the total number of internal degree of freedom of baryons. Hence, at equilibrium, the total baryon number density is read as $n_{B} = \frac{1}{6}\mu_{B}g_{b}T^{2}$. A gravitational interaction between the derivative of the Ricci scalar curvature and the baryon number current dynamically breaks CPT in an expanding universe \cite{Davo/2004}. The ratio of baryon to entropy below critical temperature $T_{D}$ is given by\\
\begin{equation}
\label{entropy}
\dfrac{n_{B}}{s}\,\simeq\,\dfrac{-15\,g_{B}\,\dot{R}}{4\,\pi^{2}\,g_{\star s}\,M_{\star}^{2}\,T_{D}^{2}}.
\end{equation}
where $g_{\star s}\,=\,\dfrac{45s}{2\,\pi^{2}\,T^{3}}$.

From equation (\ref{entropy}), it is clear that $\dot{R} = 0$ leads $\dfrac{n_{B}}{s} = 0$. So, for gravitational baryogenesis, $\dot{R}$ must be non-vanishing.\\
The Hubble's parameter is defined as
\begin{equation}
\label{H}
H\,=\,\dfrac{\dot{a}}{a}
\end{equation}
Solving equations (\ref{ef-2}), (\ref{ef-3}), (\ref{es}) and (\ref{H}) together we obtain the following  system of equations
\begin{equation}
\label{H-1}
\dot{H}\,=\,-4\,\pi\,G\,(1+\omega)\,\rho.
\end{equation}
\begin{equation}
\label{es-2}
\omega\,=\,-1-\dfrac{\dot{H}}{\dfrac{3\,H^2}{2}-\dfrac{2\,\Lambda_0^2}{3}}\,=\,-1-\dfrac{2\,\dot{H}}{3\,H^2-\Lambda}.
\end{equation}
\begin{figure}[ht!]
\centering
\includegraphics[width=10cm,height=7cm,angle=0]{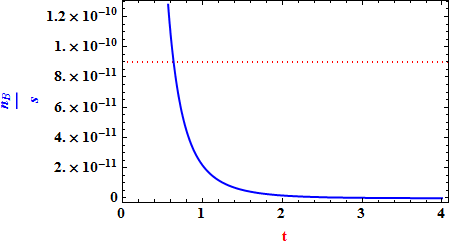}
\caption{Plot of $\dfrac{n_{B}}{s}$ versus $t$ for $g_{\star s} = 106$, $g_{B} \sim 1$, $T_{D} = 2\times 10^{12}\;Gev$ and $M_{\star} = 2\times 10^{16}\; Gev$. Blue line indicates theoretical profile while dotted red line
represents current observational constraint.}
\end{figure}
The Ricci scalar curvature for flat FLRW model is given by
\begin{equation}
\label{ricci}
R\,=\,-6\,(\dot{H}+2\,H^{2})
\end{equation}
Using equations (\ref{ef-3}) and (\ref{H-1}), equation (\ref{ricci}) leads
\begin{equation}
\label{ricci-1}
R\,=\,8\,\pi\,G\,[3\,(1+\omega)-4]\,\rho-4\,\Lambda.
\end{equation}
Therefore, the expression for Ricci scalar is read as
\begin{equation}
\label{ricci-2}
R\,=\,\dfrac{4\,\Lambda_0^2}{3}\,\Big(1+3\,\omega-2\,\cosh\left[2\,\Lambda_0\,(1+\omega)\,t+t_0\right]\Big)\,\mathrm{csch}^2\left[\Lambda_0\,(1+\omega)\,t+t_0\right].
\end{equation}
Therefore, the first derivative of $R$ is given by
\begin{equation}
\label{ricci-3}
\dot{R}\,=\,-\dfrac{8\,\Lambda_0^3}{3}\,(1+\omega)\,(3\,\omega-1)\,\coth\left[\Lambda_0\,(1+\omega)\,t+t_0\right]\,\mathrm{csch}^2\left[\Lambda_0\,(1+\omega)\,t+t_0\right].
\end{equation}
From equation (\ref{ricci-3}), we observe that $\dot{R} = 0$ for $\omega = \frac{1}{3}$ or $\omega = -1$.\\
Equations (\ref{entropy}) and (\ref{ricci-3}) lead to
\begin{equation}
\label{entropy-1}
\dfrac{n_{B}}{s}\,\simeq\,\dfrac{15\,g_{B}}{4\,\pi^{2}\,g_{\star s}\,M_{\star}^{2}\,T_{D}^{2}}\,\left[\dfrac{8\,\Lambda_0^3}{3}\,(1+\omega)\,(3\,\omega-1)\,\coth\left[\Lambda_0\,(1+\omega)\,t+t_0\right]\,\mathrm{csch}^2\left[\Lambda_0\,(1+\omega)\,t+t_0\right]\right].
\end{equation}

Figure 4 shows baryon to entropy ratio as a function of age $(t)$. We choose $g_{\star s}\,=\,106$, $g_{B}\,\sim 1$, $T_{D}\,=\,2\,\times\,10^{12}\;Gev$ and $M_{\star}\,=\,2\,\times\,10^{16}\; Gev$. This choice of $M_{\star}$ has been detected in the form of gravitational waves by LIGO and reported in Davoudiasl et al \cite{Davo/2004}. The observations indicate that $\dfrac{n_{B}}{s}\,\sim\,9.0\,\times\,10^{-11}$. Our obtained theoretical value of baryon entropy ratio for accepted range of age is in good agreement with observations.\\
\section{Conclusion}
In this paper, we have investigated an exact gravitational baryogenesis of isotropic and homogeneous universe in the framework of general relativity. We observe that the derived model presents an accelerated expansion of universe at present epoch. The energy density and pressure of the universe decreases as time increases. For baryogenesis interaction involving derivative of Ricci scalar $(\dot{R})$, our model shows the baryon entropy ratio is in good contrast with their corresponding observational value. We find that baryon to entropy ratio is proportional to $\dot{R}$ and the universe is comprising predominantly with perfect fluid having barotropic equation of state $p\,=\,\omega\,\rho$. In the expansion era, the universe was entered in accelerated phase and cosmic dynamics is governed by cosmological constant at present epoch. For cosmological constant dominated universe, $\dfrac{n_{B}}{s}$ is found to be in good agreement with its value obtained from observations.

The other physical parameters like Hubble parameter $H$, energy density $\rho$, and scale factor $a$ are depicted in Fig. 1. The Hubble parameter $H$ and energy density $\rho$ are positive and tends to null at infinite time while the scale factor is increasing function of time. The dynamics of deceleration parameter shows a signature flipping from positive to negative value which turn into imply that the current universe is in accelerating phase. Therefore, our model represents a viable model of accelerating universe in which gravitational baryogenesis occurs. Furthermore, we also note that $\dot{R} = 0$ for $\omega = \frac{1}{3}$ (radiation dominated universe) or $\omega = -1$ (pure cosmological constant universe). Thus, we conclude that a radiation dominated universe or a pure cosmological constant universe leads to no baryon asymmetry.

\section*{Acknowledgments}
The authors are grateful to the honorable referees and editor for illuminating suggestions that have significantly improved this work in terms of research quality and presentation.


\end{document}